\documentclass[fleqn,usenatbib]{mnras}

\usepackage{newtxtext,newtxmath}

\usepackage[T1]{fontenc}
\usepackage{ae,aecompl}

\usepackage{graphicx}	
\usepackage{amsmath}	
\usepackage{amssymb}	

\title[The distance and supernovae of NGC~6946]{The distance, supernova rate and supernova progenitors of NGC~6946}

\author[J.J. Eldridge \& L. Xiao]{
J.J. Eldridge,$^{1}$\thanks{E-mail: j.eldridge@auckland.ac.nz}
Lin Xiao,$^{2}$\thanks{E-mail: lxiao33@ustc.edu.cn}
\\
$^{1}$Department of Physics, Private Bag 92019, University of Auckland, Auckland 1010, New Zealand\\
$^{2}$CAS Key Laboratory for Research in Galaxies and Cosmology, Department of Astronomy, University of Science and Technology of China, Hefei, 230026, China\\
}

\date{Accepted XXX. Received YYY; in original form ZZZ}

\pubyear{2015}

\begin{document}
\label{firstpage}
\pagerange{\pageref{firstpage}--\pageref{lastpage}}
\maketitle

\begin{abstract}
The distance to the fireworks galaxy NGC 6946 is highly uncertain. Recent distance estimates using the tip of the red giant branch of 7.7 to 7.8~Mpc are larger than the distance commonly assumed for studying supernovae in this galaxy of 5.9~Mpc. Here we use the high supernova rate of the galaxy to derive the star-formation rate and predict the galaxy's FUV flux. We also account for dust extinction by different methods to derive a distance of 7.9$\pm$4.0~Mpc for NGC~6946. We then use the new distance to re-evaluate the understanding of the supernova progenitors 2002hh, 2004et, 2017eaw, the possible electron capture event 2008S and the possible black-hole forming event N6946-BH1. For the latter two exotic events the new distance improves the consistency between the observed progenitors and the stellar models that give rise to these events. From our findings we strongly recommend that all future studies of NGC~6946 must use the greater distance to the galaxy of $7.72\pm0.32$~Mpc of \citet{2018AJ....156..105A}.
\end{abstract}




\section{Introduction}

Supernovae (SNe) are rare events in our own Galaxy with only one or two expected in a century. Today most supernovae are discovered in distant galaxies and in most of these there is only one event per galaxy. There are several galaxies that have two or more observed SNe \citep{2013A&A...550A..69A}. The most extreme example is that of NGC~6946, the ``Fireworks'' galaxy. There have been 10 observed core-collapse SNe since 1917 and recently a candidate core-collapse event with no SN luminous display due to black-hole formation was discovered \citep{2013A&A...550A..69A,2017MNRAS.468.4968A,2017ATel10372....1D}. This high rate of SNe is unique and provides a novel avenue to place a constraint on the star formation rate and distance to the galaxy.

The rate of stellar death is closely linked to the rate of stellar birth \citep[e.g][]{2015MNRAS.452.2597X,2018arXiv180501213X}. The time from when a massive star is born to when it may explode in a core-collapse SNe ranges from 3Myrs to around a few 100~Myrs \citep[e.g.][]{2017A&A...601A..29Z}. Therefore the rate of observed core-collapse events can give an indication of the star-formation rate of a galaxy over the last 100~Myrs. There are uncertainties but any result is still a robust lower limit on the star-formation rate. It is a lower limit as any missed or failed supernovae would imply a higher star formation rate of a galaxy. If we use this star formation rate estimate we can then predict the expected luminosity of the galaxy and compare this to the observed (and dust corrected) luminosity to arrive at a distance estimate.

From searching the NASA Extragalactic Database (NED) we found distance estimates for NGC~6946 that range from 4.020 to 12.700~Mpc, or distance moduli from 28.02 to 30.51. The significant uncertainty in distance makes it difficult to constrain the nature of the core-collapse SN progenitors that have been observed in NGC~6946. These include the relatively normal IIP SNe 2002hh, 2004et and 2017eaw as well as the more unique events SN 2008S and N6946-BH1. Most studies concerning these SNe assume a distance calculated by \citet{2000ApJ...529..786M} of  5.9\,Mpc \citep[e.g.][]{2015PASA...32...16S,2017MNRAS.468.4968A,2018MNRAS.481.2536K,2018MNRAS.474.2116D}. However two recent measurements made by studying the tip of the red giant branch, \citet{2018AJ....156..105A} and \citet{2018ApJ...860..117M} found the significantly greater distances of 7.72$\pm$0.32\,Mpc and 7.83$\pm$0.29\,Mpc. These distanes imply that all sources in the galaxy are 0.24 dex, 70 per cent, brighter. Such a change in luminosity has significant implications for the two unusual events, 2008S and N6946-BH1, whose previous luminosity estimates were not consistent with model predictions.

In this paper we first demonstrate how the SN rate can be used to provide a distance estimate to NGC~6946. We then reanalyse the known SN progenitors in NGC~6946 with the new distance before finally summarizing our findings.

\section{Distance from the supernova rate}

In \citet{2015MNRAS.452.2597X} we investigated the link between the star-formation rate and SN rate in the galaxies of the 11HUGS survey \citep{2009ApJ...706..599L}. This was following similar studies of \citet{2012A&A...537A.132B} and \citet{2013ApJ...769..113H} but we considered the impact of interacting binaries on the nature of the stellar populations using results from BPASS v1.1 \citep{2012MNRAS.419..479E}. 

Recently we re-examined the data using the latest BPASS v2.2 models \citep{2017PASA...34...58E,2018MNRAS.479...75S} and it became apparent that one galaxy alone contributed a large number of SN to the study but had an apparently low star formation rate, this was NGC~6946. We broadened our study of this galaxy to include all the core-collapse events that have been observed within it.

For the SN rate we use the list of SN from \citet{2013A&A...550A..69A} and supplement it with SN 2017eaw \citep{2018MNRAS.481.2536K} and N6946-BH1 \citep{2017MNRAS.468.4968A}. In the 102 years up to 2018 there have been a total of 11 core-collapse events and thus a supernova rate of $0.108\pm0.033 {\rm yr^{-1}}$, assuming the errors in the counting of the number of SNe is described by a Poisson distribution.

To convert these observed values into star formation rates we take the average of the SN rate expected for 1~M$_{\odot}$yr$^{-1}$ of star formation at the metallicities representative for NGC~6946 (metallicity mass fractions, Z=0.006, 0.008, 0.010, 0.014 and 0.020). This value is 0.0090$\pm$0.0002 events yr$^{-1}$ for a star formation rate of 1~M$_{\odot}$yr$^{-1}$ from our fiducial BPASS v2.2 models. Using this value gives a star formation rate of 12.1$\pm$3.7\,M$_{\odot}$yr$^{-1}$.

From the same BPASS models we can also calculate that $\log(L_{\rm FUV}/{\rm erg \, s^{-1}} \,M^{-1}_{\odot}\,{\rm yr})$=43.37$\pm$0.03. Thus we can estimate that the FUV luminosity of NGC~6946 should be (2.8$\pm$0.9$)\times10^{44}\, {\rm erg \, s^{-1}}$.

The observed FUV luminosity of NGC~6946 is difficult to estimate. There is significant foreground extinction as well as the intrinsic extinction within the galaxy itself. One value for the FUV luminosity was determined by \citet{2009ApJ...706..599L} as $1.302\times10^{44}\, {\rm erg \, s^{-1}}$ at a distance of 5.9~Mpc, half our FUV luminosity estimated from the SN rate. This value is equivalent to an incident FUV flux of 3.13$\times$10$^{-8}\, {\rm erg \, s^{-1}\, cm^{-2}}$. This value includes a correction due to dust extinction based on the observed H$\alpha$ extinction.

To verify this luminosity we take the GALEX FUV and NUV fluxes found on NED from \citet{2009ApJ...703.1569M}. We use the relations of \citet{2011ApJ...741..124H} to estimate A(FUV) directly from the FUV-NUV colour and find a FUV flux of 4.8$\pm$1.4$\times$10$^{-8}\, {\rm erg \, s^{-1}\, cm^{-2}}$. While 53 per cent greater than the \citet{2009ApJ...706..599L} value they are a similar order of magnitude and suggest that there cannot be any more significant dust extinction.

Using the above FUV fluxes we can arrive at two distance estimates for NGC 6946. These are 8.7$\pm$2.7~Mpc for the \citet{2009ApJ...706..599L} flux and 7.0$\pm3.0$~Mpc using our flux estimate. Taking the mean we find a combined distance estimate of 7.9$\pm$4.0~Mpc with the error on this distance is dominated by the number of SN our rate is dependent on and the effect of dust extinction. These values are agreement with those found by \citet{2018AJ....156..105A} and \citet{2018ApJ...860..117M} and greater than the distances usually assumed when studying SN and their progenitors within NGC~6946. We therefore strongly recommend that all future work must use a greater distance than 5.9~Mpc with the distance of 7.72$\pm$0.32\,Mpc from \citet{2018AJ....156..105A} being the best to use.


\section{Re-evaluating the supernova progenitors in NGC~6946}

\begin{figure*}
\begin{center}
\includegraphics[width=168mm]{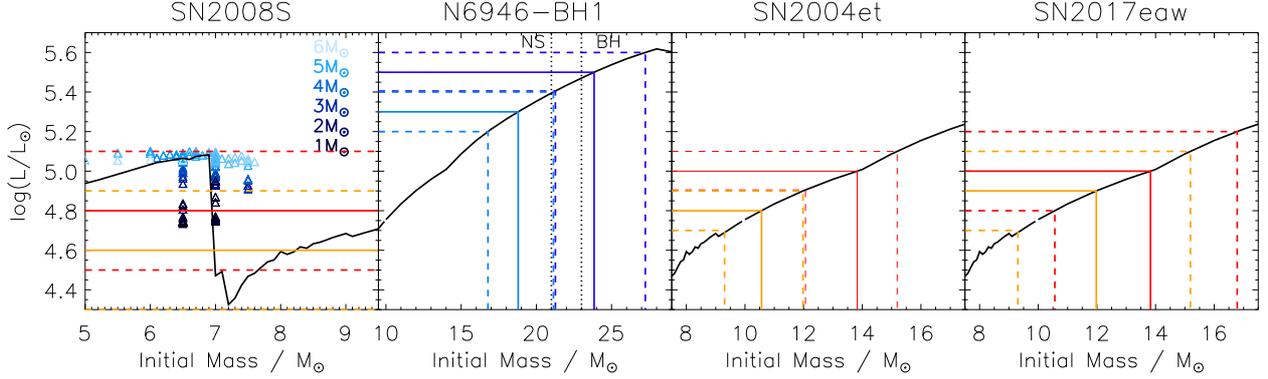}
\caption{The BPASS initial stellar mass to final pre-SN luminosity relation for single stars models at a metallicity of $Z=0.010$, suitable for NGC~6946. The four panels are for four of the different core-collapse events in NGC~6946. We do not include 2002hh as the luminosity limit is higher than all our theoretical luminosities. In the first panel for SN 2008S the first peak is due to the stars being super-AGB stars that have gone through second dredge-up, with the models beyond the rapid decrease being red supergiant progenitors. The triangles are binary models that would explode in a electron capture supernovae. The colour coding is representative of the expected ejecta mass of each binary model at the time of explosion. The remaining panels show the mass-luminosity relation for red supergiants which we use the luminosity to estimate the initial mass of the progenitor for the three observed progenitors. The coloured lines represent the luminosity estimates based on previous distance estimates (orange or light blue) or our the new distance (red or dark blue). These are then used to estimate the initial mass of the progenitor stars. The solid lines represent the best estimate while the dashed lines represent the 1$\sigma$ limits. In the second panel we include to vertical dotted lines to indicate the region where the remnants of the core collapse are neutron stars (NS) or black holes (BH) with the region between being uncertain in their nature. } 
\label{Fig1}
\end{center}
\end{figure*}

\begin{table}
	\centering
	\caption{The luminosity measurements, the distances used and the initial mass estimates in the literature. We now include our revised luminosity and initial mass estimates. References: (1) \citet{2018MNRAS.474.2116D}, (2) \citet{2015PASA...32...16S}, (3) \citet{2018MNRAS.481.2536K}, (4) \citet{2009MNRAS.398.1041B}, (5) \citep{2017MNRAS.468.4968A}.}
	\label{tab:sfr_table}
	\begin{tabular}{ccccc} 
		\hline
        	\hline
        SN & $\mu$ & $\log(L/L_{\odot})$  &  $M_{\rm i}/M_{\odot}$ & Reference \\
        \hline
2002hh     &  28.85$\pm$0.07 &  $<5.55^{+0.06}$  &  $ <26.3^{+1.6}$ & (1)   \\
           &  28.85$\pm$0.15 &  $<5.0$           &  $<15$ & (2)   \\
           &  29.43$\pm$0.09 &  $<5.78^{+0.06}$  & --  &   \\
           \hline
2004et     &  28.85$\pm$0.07 &  $4.77\pm0.07$    &  $10.7^{+0.9}_{-0.8}$ & (1)   \\
           &  28.85$\pm$0.15 &  $4.8\pm0.2$      &  $12^{+3}_{-3}$ & (2)   \\
           &  29.43$\pm$0.09 &  $5.00\pm0.1$    & $14^{+1}_{-2}$   &                              \\
           \hline
2017eaw    &  29.13$\pm$0.05 &  $4.9\pm0.2$      &  $12^{+3}_{-3}$ & (3)   \\    
           &  29.43$\pm$0.09 &  $5.0\pm0.2$     &   $14^{+3}_{-3.5}$    &          \\
           \hline
2008S      &  28.78$\pm$0.08 &  $4.6\pm0.3$      &  --   & (4) \\
           &  29.43$\pm$0.09 &  $4.86\pm0.3$    &   $7\pm1$ &     \\
           \hline
N6946  &  28.87          &  $5.3$            &  25 &  (5) \\
  -BH1         &  29.43$\pm$0.09 &  $5.5\pm0.1$     &  $24^{+3}_{-3}$                  &             \\
\hline
\hline\label{table2}
\end{tabular}
\end{table}

\subsection{SNe 2002hh, 2004et and 2017eaw}

As mentioned by \citet{2018AJ....156..105A} the greater distance to NGC~6946 implies that all supernovae are brighter, this is also true for all observed progenitor stars. We first consider the three typical type IIP supernova progenitors. For these progenitors an increased luminosity results in a larger initial mass estimates but the progenitors are still typical red supergiants.

Previous studies of SNe 2002hh and 2004et in \citet{2015PASA...32...16S} and \citet{2018MNRAS.474.2116D} assumed a distance of 5.9~Mpc. While SN 2017eaw was studied in \citet{2018MNRAS.481.2536K} assuming a distance of 6.72~Mpc. Taking a new distance of 7.72~Mpc this means for the supernovae their luminosities should increase by 0.2 and 0.1 dex respectively. We show the luminosity increases and the estimated masses in Figure \ref{Fig1} and Table \ref{table2}.

For SN~2002hh in \citet{2018MNRAS.474.2116D} a pre-explosion luminosity limit of $\log(L/L_{\odot})<5.55$ was derived which at the new distance becomes a limit of  $\log(L/L_{\odot})<5.78$. This increase means that there is now no constraint possible from this event. The limit is above the luminosity of the brightest red supergiant models we have.

The progenitor star of SN~2004et was detected with $\log(L/L_{\odot})=4.8\pm0.1$ in both previous studies which becomes 5.0$\pm$0.1 with the new distance. This changes the detected mass from around 10-11M$_{\odot}$ to $14^{+1}_{-2}$M$_{\odot}$. 

Finally the observed progenitor of SN~2017eaw with the updated distance is constrained to the same initial mass although with larger errors, $14^{+3}_{-3.5}$M$_{\odot}$.

We note that the increase of all these mass constraints of the progenitors suggests a possible solution to the red supergiant problem \citep[e.g.][]{2015PASA...32...16S,2018MNRAS.474.2116D}. This is the supposed lack of massive red supergiant progenitors above ~20M$_{\odot}$. It could also explain why so many type IIP progenitors have a luminosity near the minimum luminosity expected from stellar models \citep[e.g.][]{2011MNRAS.417.1417F}. If there is a consistent underestimate of the distance to the host galaxies this would lead to an underestimate of the masses of SN progenitors. Combined with possibly incorrect assumed bolometric corrections, as discussed by \citet{2018MNRAS.474.2116D}, we suggest it is possible that current estimates of type IIP progenitor masses may be underestimates.

\subsection{SN 2008S and N6946-BH1}

The remaining two events for which progenitor constraints exist are more unique events. For both of these events the lower luminosities from the shorter distance do not match their suspected nature when compared to BPASS models. The new distance increases the observed luminosities and these match the predicted BPASS models more closely.

SN~2008S is suspected to have been an electron capture SN, although is still debated \citep{2009MNRAS.398.1041B,2011ApJ...741...37K,2012ApJ...750...77S,2014A&A...569A..57M}. A key piece of evidence was evolution of the explosion that suggested the SN was the result of an electron-capture core-collapse of a oxygen-neon core in a super-asymptotic giant branch (SAGB) star, rather than core-collapse of an iron core of a red supergiant. The problem with this picture was that the luminosity derived for the progenitor star was too low to have been a SAGB star. SAGB stars are expected to have luminosities of $\log(L/L_{\odot})~4.9$ to 5.1 for single star models as shown in Figure \ref{Fig1}. Searching through our BPASS binary star models for electron-capture SN progenitors we found that luminosities down to $\log(L/L_{\odot})=4.75$ are possible but the majority of progenitors are above $\log(L/L_{\odot})=5$. These high luminosities are due to the temperatures around the compact oxygen-neon core being high leading to stronger hydrogen burning compared to the cores of stars that do not experience second dredge-up and remain as red supergiants \citep{2007MNRAS.376L..52E}. Stars that have experienced binary interactions have lower luminosities due to lower hydrogen abundances and lower mass envelopes reducing the hydrogen burning luminosity.

Using a distance of 5.7~Mpc the luminosity was derived to be $\log (L/L_{\odot})=4.6\pm0.3$ \citep{2009MNRAS.398.1041B}. With the greater distance the new luminosity the luminosity is closer to $\log (L/L_{\odot})=4.8\pm0.3$. This higher luminosity overlaps will all predicted electron-capture BPASS electron-capture SN models. In comparison the previous luminosity only match a tenth of our electron-capture SN models. Furthermore the models now closest to the best luminosity estimate have an ejecta mass of only 1M$_{\odot}$. Such lower ejecta masses do match with modelling of electron-capture lightcurves by \citet{2014A&A...569A..57M} as well as the fact that the progenitor was enshouded in a dusty cocoon of material that was lost close to the explosion.

If the SN was not a electron capture SN it would have been the explosion of a 11M$_{\odot}$ red supergiant. However 2008S has little in common with other such SN from progenitors in this mass range \citep{2015PASA...32...16S,2018MNRAS.474.2116D}. In conclusion using the new distance to determine the luminosity of the progenitor we find it more likely than before that SN 2008S was an electron capture event. It also was likely to have been the result of a binary interaction which was suggested by \citet{2004ApJ...612.1044P} to be the most likely channel for such core-collapse events.

While SN 2008S is at the minimum mass for a core-collapse supernovae, N6946-BH1 is at the maximum mass for a core-collapse of a red supergiant. This star was observed to rapidly fade with a lightcurve that was in agreement with a source powered by a feeding black hole \citep{2017MNRAS.468.4968A}. When the nature of the progenitor was examined this study assumed a distance of 5.95~Mpc. Therefore again this increases the suggested progenitor luminosity from $\log(L/L_{\odot})=5.3$ to 5.5. As shown in Figure \ref{Fig1} this increase in the luminosity implied a mass for the progenitor star or 24\,M$_{\odot}$ that is above the predicted mass limit for the formation of a black hole. This is also consistent with the mass estimated from the surrounding stellar population by \citet{2018ApJ...860..117M}. Again while it is only one event tension between the mass estimates for black hole formation and the luminosity of the observed progenitor stars are reduced by this study. We note that this result is in agreement with the work of \citet{2018ApJ...860..117M}.

\section{Conclusions}

We have re-evaluated the distance to NGC~6946. The value of 5.9~Mpc is incorrect and should not be used in future studies. We used the observed SN rate to estimate the star-formation rate and thus the FUV luminosity of the galaxy. Using the standard candel method to derive distances of 7.9$\pm$4.0~Mpc. This is consistent with the more recent distances estimated by \citet{2018AJ....156..105A} and \citet{2018ApJ...860..117M} of 7.72$\pm$0.32\,Mpc and 7.83$\pm$0.29\,Mpc respectively.

We strongly recommend that all future studies concerning NGC~6946 use the updated distance of 7.72$\pm$0.32\,Mpc from \citet{2018AJ....156..105A}. Their distance is consistent with the distance of \citet{2018ApJ...860..117M} but \citet{2018AJ....156..105A} took additional steps to reduce contamination and is, therefore, preferred. 

We re-evaluated the nature of the 5 progenitors of observed core collapse events in NGC~6946 using the distance of  7.72$\pm$0.32\,Mpc. We found that the mass constraints change significantly. The two most significant changes are, first, that the observed progenitor of SN 2008S is now more consistent with predictions for the progenitors of electron capture SNe. We suggest that SN 2008S was an electron capture SN with the progenitor having experienced a binary interaction before explosion. Second, the mass estimate for N6946-BH1 is closer to 24M$_{\odot}$ and agrees with the mass range for black hole formation in the progenitor star. Further detection of similar disappearing stars will further refine our understanding of the mass range at which we expect black hole formation to occur at core-collapse. Although we note that the exact mass range where stars produce black holes may be a chaotic region and not simply described by a single mass range as shown by \citet{2018ApJ...860...93S} for example. 

Finally we also suggest that the distance to other host galaxies of SN progenitors are perhaps considered to be too accurate and that any study where the result hinges sensitively on the luminosity of a source to infer its nature should not underestimate the inherent errors in distance measurements.

\section*{Acknowledgements}

JJE acknowledges travel funding and support from the University of Auckland. LX acknowledges support from the China Postdoctoral Science Foundation (grant No.2018M642524). LX acknowledge the grant from the National Key R\&D Program of China (2016YFA0400702), the National Natural Science Foundation of China (No. 11673020 and No. 11421303), and the National Thousand Young Talents Program of China. LX would also like to thank the China Scholarship Council for its funding her PhD study at the University of Auckland and the travel funding and support from the University of Auckland.




\bibliographystyle{mnras}
\bibliography{ngc6946} 

\begin{thebibliography}{}
\makeatletter
\relax
\def\mn@urlcharsother{\let\do\@makeother \do\$\do\&\do\#\do\^\do\_\do\%\do\~}
\def\mn@doi{\begingroup\mn@urlcharsother \@ifnextchar [ {\mn@doi@}
  {\mn@doi@[]}}
\def\mn@doi@[#1]#2{\def\@tempa{#1}\ifx\@tempa\@empty \href
  {http://dx.doi.org/#2} {doi:#2}\else \href {http://dx.doi.org/#2} {#1}\fi
  \endgroup}
\def\mn@eprint#1#2{\mn@eprint@#1:#2::\@nil}
\def\mn@eprint@arXiv#1{\href {http://arxiv.org/abs/#1} {{\tt arXiv:#1}}}
\def\mn@eprint@dblp#1{\href {http://dblp.uni-trier.de/rec/bibtex/#1.xml}
  {dblp:#1}}
\def\mn@eprint@#1:#2:#3:#4\@nil{\def\@tempa {#1}\def\@tempb {#2}\def\@tempc
  {#3}\ifx \@tempc \@empty \let \@tempc \@tempb \let \@tempb \@tempa \fi \ifx
  \@tempb \@empty \def\@tempb {arXiv}\fi \@ifundefined
  {mn@eprint@\@tempb}{\@tempb:\@tempc}{\expandafter \expandafter \csname
  mn@eprint@\@tempb\endcsname \expandafter{\@tempc}}}

\bibitem[\protect\citeauthoryear{{Adams}, {Kochanek}, {Gerke}, {Stanek}  \&
  {Dai}}{{Adams} et~al.}{2017}]{2017MNRAS.468.4968A}
{Adams} S.~M.,  {Kochanek} C.~S.,  {Gerke} J.~R.,  {Stanek} K.~Z.,   {Dai} X.,
  2017, \mn@doi [\mnras] {10.1093/mnras/stx816}, \href
  {http://adsabs.harvard.edu/abs/2017MNRAS.468.4968A} {468, 4968}

\bibitem[\protect\citeauthoryear{{Anand}, {Rizzi}  \& {Tully}}{{Anand}
  et~al.}{2018}]{2018AJ....156..105A}
{Anand} G.~S.,  {Rizzi} L.,   {Tully} R.~B.,  2018, \mn@doi [\aj]
  {10.3847/1538-3881/aad3b2}, \href
  {http://adsabs.harvard.edu/abs/2018AJ....156..105A} {156, 105}

\bibitem[\protect\citeauthoryear{{Anderson} \& {Soto}}{{Anderson} \&
  {Soto}}{2013}]{2013A&A...550A..69A}
{Anderson} J.~P.,  {Soto} M.,  2013, \mn@doi [\aap]
  {10.1051/0004-6361/201220600}, \href
  {http://adsabs.harvard.edu/abs/2013A%26A...550A..69A} {550, A69}

\bibitem[\protect\citeauthoryear{{Botticella} et~al.,}{{Botticella}
  et~al.}{2009}]{2009MNRAS.398.1041B}
{Botticella} M.~T.,  et~al., 2009, \mn@doi [\mnras]
  {10.1111/j.1365-2966.2009.15082.x}, \href
  {http://adsabs.harvard.edu/abs/2009MNRAS.398.1041B} {398, 1041}

\bibitem[\protect\citeauthoryear{{Botticella}, {Smartt}, {Kennicutt},
  {Cappellaro}, {Sereno}  \& {Lee}}{{Botticella}
  et~al.}{2012}]{2012A&A...537A.132B}
{Botticella} M.~T.,  {Smartt} S.~J.,  {Kennicutt} R.~C.,  {Cappellaro} E.,
  {Sereno} M.,   {Lee} J.~C.,  2012, \mn@doi [\aap]
  {10.1051/0004-6361/201117343}, \href
  {http://adsabs.harvard.edu/abs/2012A%26A...537A.132B} {537, A132}

\bibitem[\protect\citeauthoryear{{Davies} \& {Beasor}}{{Davies} \&
  {Beasor}}{2018}]{2018MNRAS.474.2116D}
{Davies} B.,  {Beasor} E.~R.,  2018, \mn@doi [\mnras] {10.1093/mnras/stx2734},
  \href {http://adsabs.harvard.edu/abs/2018MNRAS.474.2116D} {474, 2116}

\bibitem[\protect\citeauthoryear{{Dong} \& {Stanek}}{{Dong} \&
  {Stanek}}{2017}]{2017ATel10372....1D}
{Dong} S.,  {Stanek} K.~Z.,  2017, The Astronomer's Telegram, \href
  {http://adsabs.harvard.edu/abs/2017ATel10372....1D} {10372}

\bibitem[\protect\citeauthoryear{{Eldridge} \& {Stanway}}{{Eldridge} \&
  {Stanway}}{2012}]{2012MNRAS.419..479E}
{Eldridge} J.~J.,  {Stanway} E.~R.,  2012, \mn@doi [\mnras]
  {10.1111/j.1365-2966.2011.19713.x}, \href
  {http://adsabs.harvard.edu/abs/2012MNRAS.419..479E} {419, 479}

\bibitem[\protect\citeauthoryear{{Eldridge}, {Mattila}  \& {Smartt}}{{Eldridge}
  et~al.}{2007}]{2007MNRAS.376L..52E}
{Eldridge} J.~J.,  {Mattila} S.,   {Smartt} S.~J.,  2007, \mn@doi [\mnras]
  {10.1111/j.1745-3933.2007.00285.x}, \href
  {http://adsabs.harvard.edu/abs/2007MNRAS.376L..52E} {376, L52}

\bibitem[\protect\citeauthoryear{{Eldridge}, {Stanway}, {Xiao}, {McClelland},
  {Taylor}, {Ng}, {Greis}  \& {Bray}}{{Eldridge}
  et~al.}{2017}]{2017PASA...34...58E}
{Eldridge} J.~J.,  {Stanway} E.~R.,  {Xiao} L.,  {McClelland} L.~A.~S.,
  {Taylor} G.,  {Ng} M.,  {Greis} S.~M.~L.,   {Bray} J.~C.,  2017, \mn@doi
  [\pasa] {10.1017/pasa.2017.51}, \href
  {http://adsabs.harvard.edu/abs/2017PASA...34...58E} {34, e058}

\bibitem[\protect\citeauthoryear{{Fraser} et~al.,}{{Fraser}
  et~al.}{2011}]{2011MNRAS.417.1417F}
{Fraser} M.,  et~al., 2011, \mn@doi [\mnras]
  {10.1111/j.1365-2966.2011.19370.x}, \href
  {http://adsabs.harvard.edu/abs/2011MNRAS.417.1417F} {417, 1417}

\bibitem[\protect\citeauthoryear{{Hao}, {Kennicutt}, {Johnson}, {Calzetti},
  {Dale}  \& {Moustakas}}{{Hao} et~al.}{2011}]{2011ApJ...741..124H}
{Hao} C.-N.,  {Kennicutt} R.~C.,  {Johnson} B.~D.,  {Calzetti} D.,  {Dale}
  D.~A.,   {Moustakas} J.,  2011, \mn@doi [\apj] {10.1088/0004-637X/741/2/124},
  \href {http://adsabs.harvard.edu/abs/2011ApJ...741..124H} {741, 124}

\bibitem[\protect\citeauthoryear{{Horiuchi}, {Beacom}, {Bothwell}  \&
  {Thompson}}{{Horiuchi} et~al.}{2013}]{2013ApJ...769..113H}
{Horiuchi} S.,  {Beacom} J.~F.,  {Bothwell} M.~S.,   {Thompson} T.~A.,  2013,
  \mn@doi [\apj] {10.1088/0004-637X/769/2/113}, \href
  {http://adsabs.harvard.edu/abs/2013ApJ...769..113H} {769, 113}

\bibitem[\protect\citeauthoryear{{Kilpatrick} \& {Foley}}{{Kilpatrick} \&
  {Foley}}{2018}]{2018MNRAS.481.2536K}
{Kilpatrick} C.~D.,  {Foley} R.~J.,  2018, \mn@doi [\mnras]
  {10.1093/mnras/sty2435}, \href
  {http://adsabs.harvard.edu/abs/2018MNRAS.481.2536K} {481, 2536}

\bibitem[\protect\citeauthoryear{{Kochanek}}{{Kochanek}}{2011}]{2011ApJ...741...37K}
{Kochanek} C.~S.,  2011, \mn@doi [\apj] {10.1088/0004-637X/741/1/37}, \href
  {http://adsabs.harvard.edu/abs/2011ApJ...741...37K} {741, 37}

\bibitem[\protect\citeauthoryear{{Lee} et~al.,}{{Lee}
  et~al.}{2009}]{2009ApJ...706..599L}
{Lee} J.~C.,  et~al., 2009, \mn@doi [\apj] {10.1088/0004-637X/706/1/599}, \href
  {http://adsabs.harvard.edu/abs/2009ApJ...706..599L} {706, 599}

\bibitem[\protect\citeauthoryear{{Moriya}, {Tominaga}, {Langer}, {Nomoto},
  {Blinnikov}  \& {Sorokina}}{{Moriya} et~al.}{2014}]{2014A&A...569A..57M}
{Moriya} T.~J.,  {Tominaga} N.,  {Langer} N.,  {Nomoto} K.,  {Blinnikov} S.~I.,
    {Sorokina} E.~I.,  2014, \mn@doi [\aap] {10.1051/0004-6361/201424264},
  \href {http://adsabs.harvard.edu/abs/2014A%26A...569A..57M} {569, A57}

\bibitem[\protect\citeauthoryear{{Mould} et~al.,}{{Mould}
  et~al.}{2000}]{2000ApJ...529..786M}
{Mould} J.~R.,  et~al., 2000, \mn@doi [\apj] {10.1086/308304}, \href
  {https://ui.adsabs.harvard.edu/\#abs/2000ApJ...529..786M} {529, 786}

\bibitem[\protect\citeauthoryear{{Mu{\~n}oz-Mateos} et~al.,}{{Mu{\~n}oz-Mateos}
  et~al.}{2009}]{2009ApJ...703.1569M}
{Mu{\~n}oz-Mateos} J.~C.,  et~al., 2009, \mn@doi [\apj]
  {10.1088/0004-637X/703/2/1569}, \href
  {https://ui.adsabs.harvard.edu/\#abs/2009ApJ...703.1569M} {703, 1569}

\bibitem[\protect\citeauthoryear{{Murphy}, {Khan}, {Williams}, {Dolphin},
  {Dalcanton}  \& {D{\'{\i}}az-Rodr{\'{\i}}guez}}{{Murphy}
  et~al.}{2018}]{2018ApJ...860..117M}
{Murphy} J.~W.,  {Khan} R.,  {Williams} B.,  {Dolphin} A.~E.,  {Dalcanton} J.,
   {D{\'{\i}}az-Rodr{\'{\i}}guez} M.,  2018, \mn@doi [\apj]
  {10.3847/1538-4357/aac2be}, \href
  {http://adsabs.harvard.edu/abs/2018ApJ...860..117M} {860, 117}

\bibitem[\protect\citeauthoryear{{Podsiadlowski}, {Langer}, {Poelarends},
  {Rappaport}, {Heger}  \& {Pfahl}}{{Podsiadlowski}
  et~al.}{2004}]{2004ApJ...612.1044P}
{Podsiadlowski} P.,  {Langer} N.,  {Poelarends} A.~J.~T.,  {Rappaport} S.,
  {Heger} A.,   {Pfahl} E.,  2004, \mn@doi [\apj] {10.1086/421713}, \href
  {http://adsabs.harvard.edu/abs/2004ApJ...612.1044P} {612, 1044}

\bibitem[\protect\citeauthoryear{{Smartt}}{{Smartt}}{2015}]{2015PASA...32...16S}
{Smartt} S.~J.,  2015, \mn@doi [\pasa] {10.1017/pasa.2015.17}, \href
  {http://adsabs.harvard.edu/abs/2015PASA...32...16S} {32, e016}

\bibitem[\protect\citeauthoryear{{Stanway} \& {Eldridge}}{{Stanway} \&
  {Eldridge}}{2018}]{2018MNRAS.479...75S}
{Stanway} E.~R.,  {Eldridge} J.~J.,  2018, \mn@doi [\mnras]
  {10.1093/mnras/sty1353}, \href
  {http://adsabs.harvard.edu/abs/2018MNRAS.479...75S} {479, 75}

\bibitem[\protect\citeauthoryear{{Sukhbold}, {Woosley}  \& {Heger}}{{Sukhbold}
  et~al.}{2018}]{2018ApJ...860...93S}
{Sukhbold} T.,  {Woosley} S.~E.,   {Heger} A.,  2018, \mn@doi [\apj]
  {10.3847/1538-4357/aac2da}, \href
  {http://adsabs.harvard.edu/abs/2018ApJ...860...93S} {860, 93}

\bibitem[\protect\citeauthoryear{{Szczygie{\l}}, {Prieto}, {Kochanek},
  {Stanek}, {Thompson}, {Beacom}, {Garnavich}  \& {Woodward}}{{Szczygie{\l}}
  et~al.}{2012}]{2012ApJ...750...77S}
{Szczygie{\l}} D.~M.,  {Prieto} J.~L.,  {Kochanek} C.~S.,  {Stanek} K.~Z.,
  {Thompson} T.~A.,  {Beacom} J.~F.,  {Garnavich} P.~M.,   {Woodward} C.~E.,
  2012, \mn@doi [\apj] {10.1088/0004-637X/750/1/77}, \href
  {http://adsabs.harvard.edu/abs/2012ApJ...750...77S} {750, 77}

\bibitem[\protect\citeauthoryear{{Xiao} \& {Eldridge}}{{Xiao} \&
  {Eldridge}}{2015}]{2015MNRAS.452.2597X}
{Xiao} L.,  {Eldridge} J.~J.,  2015, \mn@doi [\mnras] {10.1093/mnras/stv1425},
  \href {http://adsabs.harvard.edu/abs/2015MNRAS.452.2597X} {452, 2597}

\bibitem[\protect\citeauthoryear{{Xiao}, {Galbany}, {Eldridge}  \&
  {Stanway}}{{Xiao} et~al.}{2018}]{2018arXiv180501213X}
{Xiao} L.,  {Galbany} L.,  {Eldridge} J.~J.,   {Stanway} E.~R.,  2018,
  preprint, \href {http://adsabs.harvard.edu/abs/2018arXiv180501213X} {}
  (\mn@eprint {arXiv} {1805.01213})

\bibitem[\protect\citeauthoryear{{Zapartas} et~al.,}{{Zapartas}
  et~al.}{2017}]{2017A&A...601A..29Z}
{Zapartas} E.,  et~al., 2017, \mn@doi [\aap] {10.1051/0004-6361/201629685},
  \href {http://adsabs.harvard.edu/abs/2017A%26A...601A..29Z} {601, A29}

\makeatother
\end{thebibliography}



\bsp	
\label{lastpage}
\end{document}